\documentclass[11pt]{article}
\usepackage{lmodern}
\usepackage[T1]{fontenc}
\usepackage[utf8]{inputenc}
\usepackage[margin=1in]{geometry}
\usepackage{amsmath,amssymb,amsthm}
\usepackage{graphicx}
\usepackage{tabularx}
\usepackage[hidelinks]{hyperref}
\usepackage{microtype}
\usepackage[font=small,labelformat=empty]{caption}
\title{An Exact Noise Floor for Contingency-Table Effect Sizes:\\ A Per-Table Reporting Gate, and How Often It Would Change Reported Magnitudes}
\author{William J. Dwyer, MD, MPH, FAAP\\ Department of Mathematics and Statistics,\\ University of Massachusetts Lowell, Lowell, MA, USA\\ \texttt{wjdwyer@trialdesign.com} \\ ORCID 0009-0004-0855-7222}
\date{}
\begin{document}
\maketitle

\begin{abstract}
Cramer's V and the other chi-square-family effect sizes are positive under exact independence, so a reported value can be smaller than what the same statistic produces on a table with no association at all. The upward bias is classical (Tschuprow 1925; Bartlett 1937; corrected by Bergsma 2013), and reporting the boundary value, the critical effect size, beside the observed one has been recommended in general form (Perugini et al. 2025); we claim neither. Our contribution is two things a general recommendation does not supply. First, an exact per-table reporting gate: the noise floor $V_{0}.95$, the value the effect size reaches by chance 5\% of the time under independence, computed from the exact margin-conditional law rather than the asymptotic critical value. The asymptotic floor miscalibrates where reporting is riskiest: on heterogeneous tables its false-alarm rate reaches 0.10 against a 0.05 target, while the exact floor holds nominal. Second, a measurement: across 4,129 real two-way tables from 291 public datasets, 21.8\% of labeled effects (cluster-robust 95\% CI 17 to 27\%) sit on tables with no significant association. We recommend reporting $V_{0}.95$, in exact form on sparse or heterogeneous tables, beside every association measure and at the design stage.
\end{abstract}


\section{Introduction}

A researcher cross-tabulates two categorical variables, runs a chi-square test, and reports an effect size, usually Cramer's V or Cohen's w, read against Cohen's thresholds as a small, medium, or large effect. Those words travel into abstracts, meta-analyses, and the power calculation for the next study, where they do real damage: pooling cannot rescue a biased magnitude, and across simulated meta-analyses the naive estimator's chance of landing within 0.01 of the truth falls from 0.15 to 0.00 as studies accumulate (companion, Dwyer 2026b). This paper is about how often those words describe nothing, and about a device that makes the problem visible at the moment of reporting.

We claim almost none of the underlying theory, and say so throughout. The upward null bias of these coefficients is classical; the idea of reporting the critical effect size beside the observed one is not new either. What is new is narrower and, we think, useful, and it is two things.

The first is an \textbf{exact per-table reporting gate}. For a table of given shape and sample size, the \textbf{noise floor $V_{0}.95$} is the value the reported statistic attains by chance alone 5\% of the time when the true association is exactly zero. Equivalently it is the chi-square critical value mapped through the same monotone function that produces the effect size, so it is the per-table critical effect size, or minimum detectable effect, that power analysis already defines but reporting never prints. A general recommendation to report that boundary computes it from the \textbf{asymptotic} critical value; the contribution here is to compute it from the \textbf{exact margin-conditional} law, because on the sparse and heterogeneous tables where reporting is riskiest the asymptotic value is itself miscalibrated (Section 3, Figure 2). The exact engine is the companion's (Dwyer 2026a); the per-table gate on the effect-size scale, and the demonstration that it is needed, are the contribution here.

The second is a \textbf{measurement}: applying the floor to 4,129 real two-way tables from 291 public datasets, 21.8\% of labeled effects sit on non-significant tables (cluster-robust 95\% CI 17 to 27\%), a figure we read as a property of this corpus rather than a survey of practice (Section 6).

Both contributions are self-contained. Everything reported here computes from the table in front of the reader: the asymptotic floor is elementary, and the exact-moment floor this paper ships is computed from the table's own margins, so the paper can be read, judged, and used without any companion. The companion papers (the point estimate, the confidence interval, and the exact heterogeneous-margin engine) are refinements and extensions a reader may take or leave; they sharpen the floor in the sparse, strongly heterogeneous corner and answer questions this paper deliberately sets aside, but the reporting gate and the measurement stand on their own.

\subsection{The theory we build on, and claim none of}

Every member of the chi-square family is a monotone increasing function of Pearson's $X^{2}$, which is non-negative with probability one, so its null expectation is strictly positive and every effect size built on it inherits that positive null expectation. The exact statement, $E[\phi ^{2}]$ = dof/(N-1) with $\phi ^{2}$ = $X^{2}/N$ and dof = (R-1)(C-1), was conjectured by Tschuprow (1925) and proven by Bartlett (1937) under multinomial sampling. Bergsma (2013) supplies the correction $\phi ^{2}$ - dof/(N-1), implemented in standard software, and notes in passing that the corrected quantity must be allowed to go negative "or else it could not be unbiased when $\phi ^{2}$ = 0", the observation that no non-negative estimator of V can be unbiased at the null. We reproduce these results, verify them numerically, and claim none of them; they are the premise of this paper.

\subsection{Why it has not changed practice}

If the theory has been settled for nearly a century, the interesting question is why it has not reached practice. Our answer is that the correction is expressed in units nobody reads. A researcher is told the critical value of $X^{2}$ on 36 degrees of freedom is 51.0; this means nothing to them and it is not what they report. They report V = 0.19 and the word "small-to-moderate." Nothing in the output tells them that on a table of that shape and size, V reaches 0.226 by chance alone. The number that would settle the question exists, is elementary, and is never shown, and on the tables where it is most needed the elementary version of it is also wrong.

\section{The family is one object}

With row sums $r_{i}$, column sums $c_{j}$, total N, expectations $E_{ij} = r_{i} c_{j} / N$, $\phi ^{2}$ = $X^{2}/N$, dof = (R-1)(C-1), and k = min(R-1, C-1), every standard chi-square-family effect size is a function of $\phi ^{2}$ alone:

Cohen's w = $\sqrt{\phi ^{2}}$; phi (2x2) = $\sqrt{\phi ^{2}}$; Cramer's V = sqrt($\phi ^{2}$ / k); Tschuprow's T = sqrt($\phi ^{2}$ / sqrt(dof)); Pearson's C = sqrt($\phi ^{2}$ / (1 + $\phi ^{2}$))

They differ only in how they rescale one quantity, so one statement corrects all five, and one floor gates all five. On a table with no association whatsoever, an ordinary 5 x 10 at N = 250 reports an expected Cramer's V of 0.189 and an expected Cohen's w of 0.377 (Figure 1), which Cohen's conventions read as small-to-moderate and medium; Cohen's w is the effect size his own power tables are indexed by, so the defect propagates into study design and not only into reporting. Figure 1 shows the whole family inflating together under a true null; the single result $E[\phi ^{2}]$ = dof/(N-1) predicts every curve.

\begin{figure}[htbp]\centering
\includegraphics[width=\linewidth]{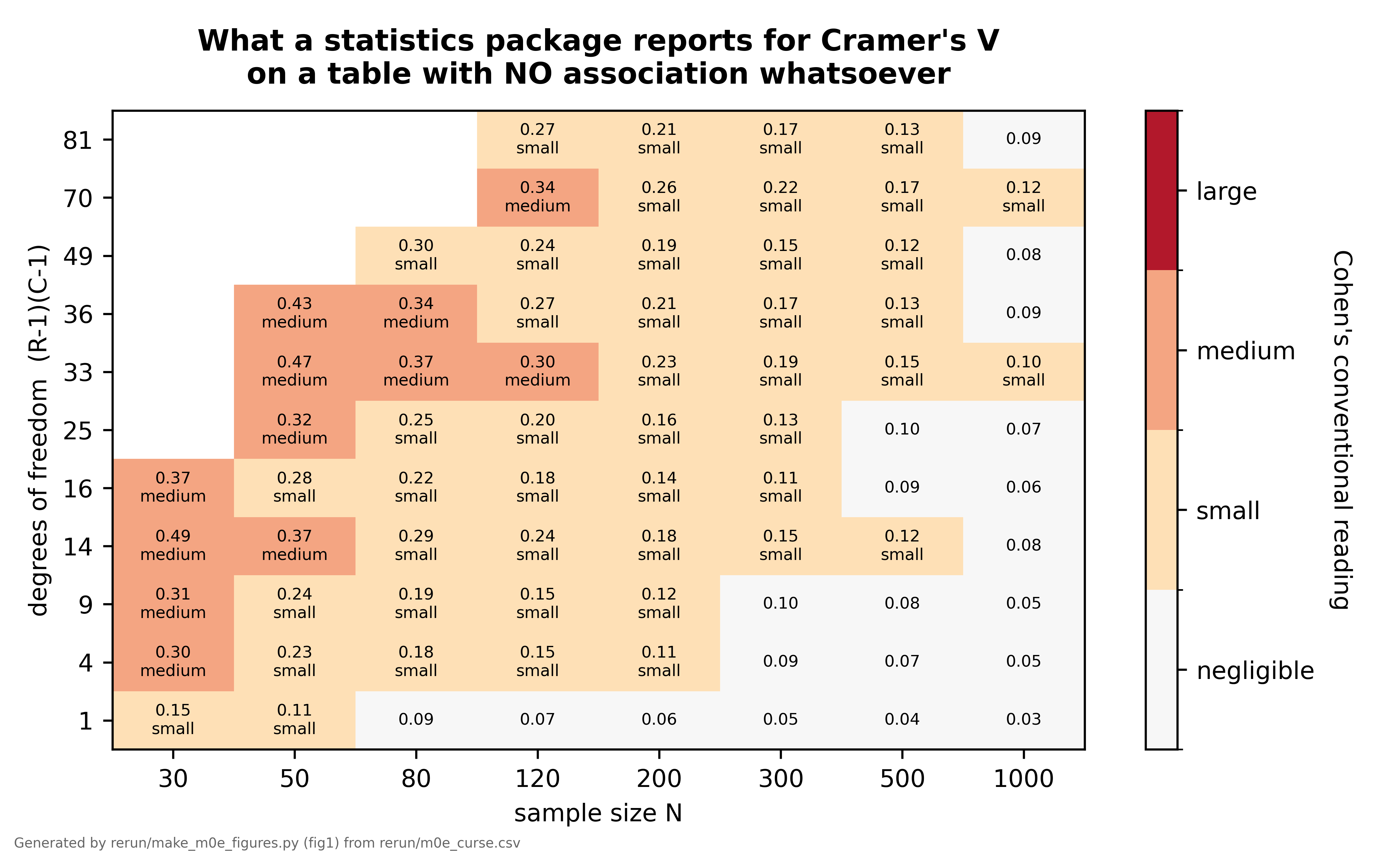}
\caption{\textbf{Figure 1.} What a statistics package reports for Cramer's V on a table with no association whatsoever, across degrees of freedom and sample size, with Cohen's conventional labels overlaid. Every chi-square-family effect size inflates the same way, and the single result $E[\phi ^{2}]$ = dof/(N-1), proven by Bartlett (1937), predicts it. Generated by rerun/make\_m0e\_figures.py; closed form, no simulation.}
\end{figure}

\section{The noise floor, and why it must be exact}

\textbf{The noise floor. For a two-way table of given margins and total N, the noise floor $V_{0}.95$ is the value the reported statistic attains by chance alone, 5\% of the time, when the true association is exactly zero. Equivalently it is the chi-square critical value, mapped through the same monotone function that produces the effect size.}

Three things must be said plainly about $V_{0}.95$: what it is, why the exact version is the point, and where even the exact version stops.

\textbf{It is the critical value, in the reader's units.} Because V is monotone in $X^{2}$, "V exceeds $V_{0}.95$" and "the chi-square test rejects" are the same event; this is not new inference. Quantitatively it is the critical effect size, or minimum detectable effect, of a standard power calculation (Cohen 1988; Bloom 1995), whose reporting beside the observed effect has recently been advocated in general form (Perugini et al. 2025). Reported as "chi-square critical value 51.0" the fact is inert; reported as "on your design, V reaches 0.226 by chance alone" it settles the question in the reader's own vocabulary, next to the number they are about to call a medium effect. Because the reported effect size and the p-value are a one-to-one transform (Hoenig and Heisey 2001), "below floor" and "not significant" coincide, and they do on 96.8\% of the 4,129 real tables below.

\textbf{The exact floor is the point, not a refinement.} A general recommendation to report the critical effect size uses the asymptotic value, $V_{0}.95$ = sqrt(chi2\_crit(0.95, dof) / (N k)). On dense balanced tables that value is right. On sparse and heterogeneous tables it is not, because the true law of the statistic is a lattice of atoms and the 0.95 quantile of that atomic law is generally not the 0.95 quantile of any continuous curve fitted to it; a three-moment chi-square fit can miss the exact tail by as much as 0.296 in probability (companion engine, Dwyer 2026a). Figure 2 makes the consequence concrete. "V exceeds the asymptotic floor" is the same event as "the Pearson chi-square rejects," so the asymptotic floor's realized false-alarm rate is the Pearson Type-I size, and over the plane of marginal heterogeneity by average expected count it runs from 0.038 in the sparse balanced corner to 0.098 where margins are heterogeneous, against a 0.05 target; the exact margin-conditional floor holds between 0.033 and 0.051 across the same plane. The exact floor is therefore not a refinement but the reason a contingency-table floor must be computed separately from the textbook value, and it is what makes this a contingency-table contribution rather than an instance of a general rule.

\begin{figure}[htbp]\centering
\includegraphics[width=\linewidth]{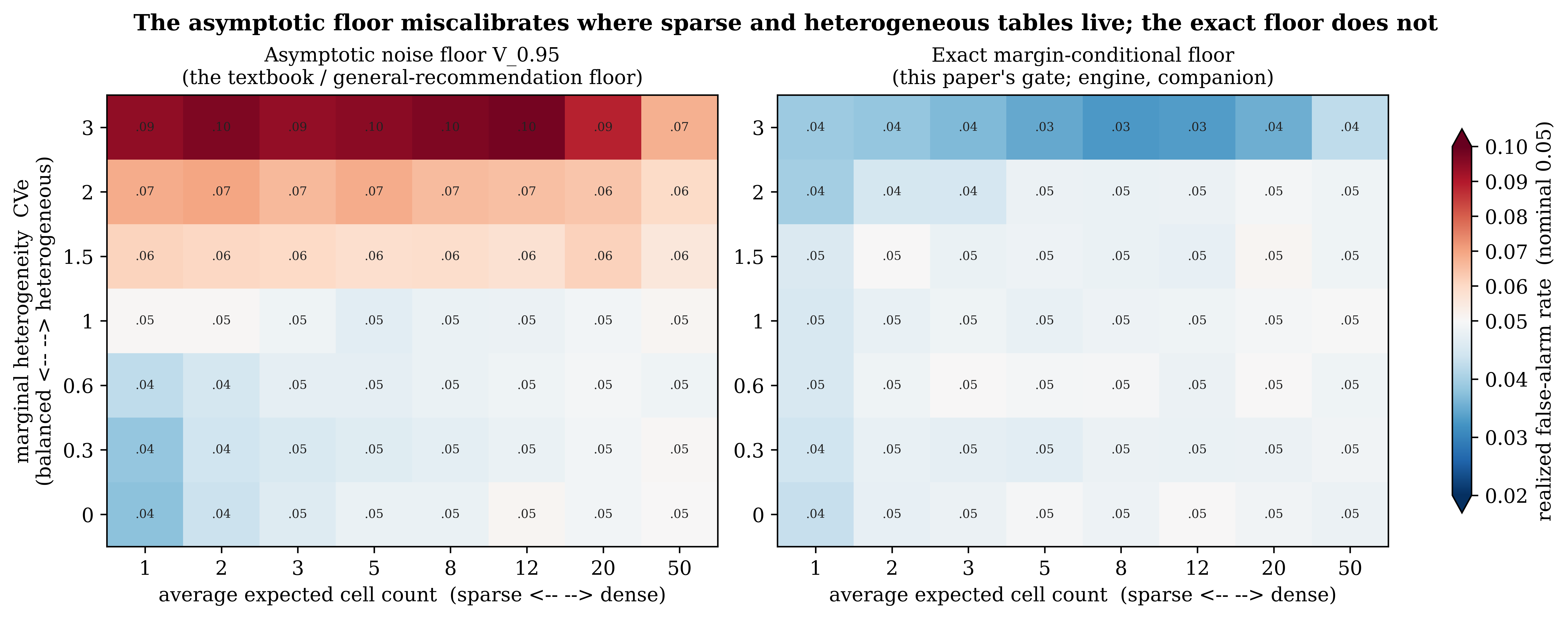}
\caption{\textbf{Figure 2.} The asymptotic noise floor $V_{0}.95$ miscalibrates where sparse and heterogeneous tables live; the exact margin-conditional floor does not. Realized false-alarm rate over the plane of marginal heterogeneity (CVe) and average expected cell count, from the deposited calibration\_plane.csv. Left: the asymptotic floor (equivalently, the Pearson chi-square), reaching 0.098 in the heterogeneous corner. Right: the exact conditional floor, holding near 0.05. Diverging colormap centered at the nominal 0.05. Generated by rerun/make\_m0e\_floor\_calibration.py.}
\end{figure}

\textbf{Where even the exact floor stops, and the design use.} The exact-moment computation shipped with this paper was calibrated on balanced margins; under strongly heterogeneous margins with a minimum expected count below about 0.05 its own false-alarm rate ranges from 0.016 to 0.066, and in that corner the exact margin-conditional engine of the companion (Dwyer 2026a) is what carries the floor. $V_{0}.95$ depends only on shape and N, so it can be computed before a single observation is collected: a 5 x 10 study planned at N = 250 has a planned-margin $V_{0}.95$ = 0.226, and any effect below 0.226 that study could report was never going to mean anything. That belongs in a protocol beside the power calculation. The floor is not a smallest effect size of interest: equivalence testing fixes a substantive threshold the analyst judges worth detecting (Lakens et al. 2018), whereas $V_{0}.95$ is the value the statistic reaches by chance under the null, a property of the design and not a judgment about the science.

\textbf{The floor is the report; the estimate and interval are the companion's.} The honest primary report is the floor, a property of shape and N alone, because no unbiased estimator of the effect size exists at any sample size (companion, Dwyer 2026b): the point estimate above the floor is a question of where to place an irreducible bias, and its treatment, together with the interval, belongs to that companion. Here the floor is the decision. Under the null with balanced margins a reported effect exceeds $V_{0}.95$ with probability 0.0500 (plus or minus 0.0006 at 500,000 replications), a one-sided statement with exactly the right error rate, and it is the statement every claim in this paper rests on.

\section{Worked examples}

The examples below are all real, all report a medium effect by Cohen's conventions, all fall below their own noise floor $V_{0}.95$, and in all of them both an approximate test and an exact permutation test find no association. We give several rather than one, deliberately: a single striking example is a single point of failure, and an earlier draft of this paper was betrayed by exactly that.

\begin{table}[htbp]\centering\small
\caption{\textbf{Table 1.} Reported effect, and what is there. Values computed by the deposited script effect\_sizes.py from rerun/m0e\_worked\_examples.csv.}
\resizebox{\textwidth}{!}{\begin{tabular}{lcccccccc}
\hline
dataset & shape & N & Cramer's V & Cohen & w & noise floor $V_{0}.95$ & p & p (exact) \\
\hline
Dactyl (fingerprint ridges) & 4x15 & 131 & 0.267 & medium & 0.462 & 0.386 & 0.92 & 0.96 \\
plantTraits (flowering x pollination) & 9x5 & 136 & 0.191 & medium & 0.381 & 0.307 & 0.83 & 0.95 \\
aldh2 (two marker loci) & 8x9 & 263 & 0.147 & medium & 0.390 & 0.227 & 0.75 & 0.86 \\
cgd (center x infection sequence) & 13x8 & 203 & 0.184 & medium & 0.487 & 0.309 & 0.85 & 0.99 \\
\hline
\end{tabular}}
\end{table}

In every row the bias-corrected V is exactly 0.000. In aldh2 the null is a substantive scientific hypothesis (independence of two marker loci is linkage equilibrium), and both tests are consistent with it. Two rows, aldh2 and cgd, have a minimum expected count of 0.02, inside the range where our own screen routes a table to an exact test; we retain them because the exact test is precisely what we ran, and it agrees. Dactyl has a minimum expected count of 0.87 and needs no such caveat.

\begin{figure}[htbp]\centering
\includegraphics[width=\linewidth]{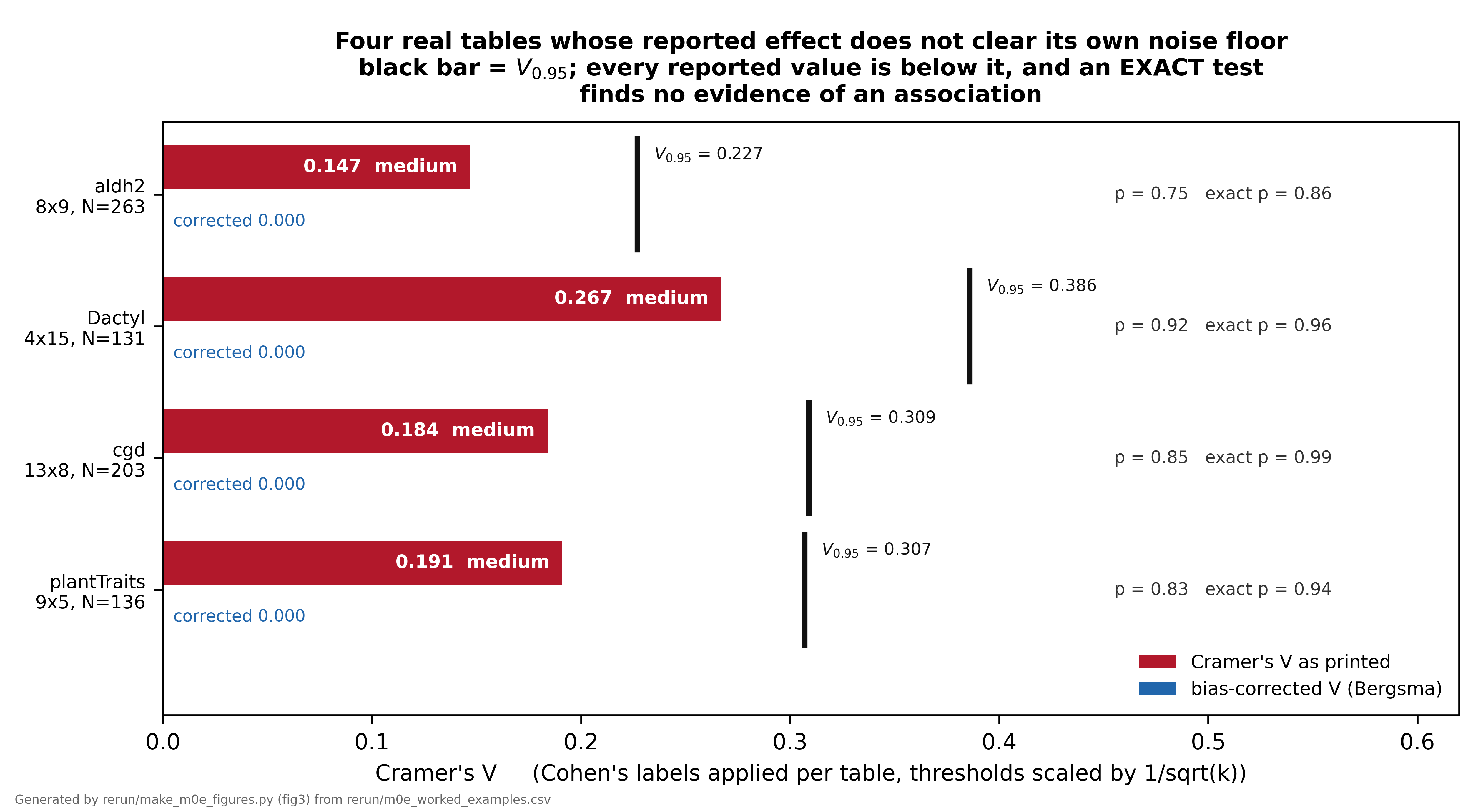}
\caption{\textbf{Figure 3.} Four real tables whose reported effect does not clear its own noise floor. The black bar marks each table's $V_{0}.95$; every reported value falls below it, and an exact permutation test finds no evidence of an association in any of them. We do not assert that these tables contain no association; we assert that the reported magnitude is not supportable. Generated by rerun/make\_m0e\_figures.py from rerun/m0e\_worked\_examples.csv.}
\end{figure}

\section{The winner's curse}

The inflation does not stop at null tables, and the mechanism compounds. The reported effect size and the p-value are monotone functions of the same $X^{2}$, so conditioning on significance is conditioning directly on a large effect size, acting on an estimate that was already biased upward.

\begin{table}[htbp]\centering\small
\caption{\textbf{Table 2.} Inflation of a real effect (designs meeting the five-expected-per-cell adequacy rule; 20,000 replications per cell). Values computed by the deposited script m0e\_winners\_curse.py.}
\resizebox{\textwidth}{!}{\begin{tabular}{lcccccc}
\hline
true V & designs & mean dof & reported V & inflation & reported V given p < .05 & inflation given p < .05 \\
\hline
0.05 & 53 & 18.1 & 0.128 & 2.6x & 0.194 & 3.9x \\
0.10 & 58 & 22.6 & 0.155 & 1.6x & 0.200 & 2.0x \\
0.20 & 46 & 11.4 & 0.231 & 1.2x & 0.259 & 1.3x \\
0.30 & 30 & 5.2 & 0.321 & 1.1x & 0.341 & 1.1x \\
\hline
\end{tabular}}
\end{table}

The restriction matters and we state it: the effect-injection procedure cannot reach large V in large tables, so an unrestricted pooling would compare different shapes across rows and would report larger inflation (3.8x and 5.2x at V = 0.05). The five-per-cell restriction removes that confound. The extreme individual cell is a 3 x 8 at N = 30, where a true V of 0.10 is reported at 6.6 times its size among significant tables, but that design has 3.9\% power; we report it as a limiting case, not a headline. Inflation is largest for the smallest true effects and the smallest studies, which is the opposite of what a reader assumes when pooling published magnitudes.

The inflation need not be simulated; it is an exact quantity with a single governing variable. Because the reported effect size and the p-value are monotone functions of the same $X^{2}$, and $X^{2}$ under an alternative is noncentral chi-square with noncentrality lambda = N k $V^{2}$, the significance-conditional effect size is a truncated moment, E[V given p < .05] = E[ $\sqrt{X^{2}}$ $1{X^{2} > c}$ ] / ( P(sig) sqrt(N k) ) with c = chi2 quantile at .95 on dof degrees of freedom, a one-dimensional integral. This exact computation reproduces the simulation of Table 2 to about 0.005 across shapes and sample sizes, and it exposes the structure: the inflation E[V given sig] / V is a function of the degrees of freedom and lambda alone, so the entire shape-by-sample-size-by-effect grid collapses onto one family of curves indexed by the degrees of freedom (Figure 4). The design consequence is a formula rather than a table: for a planned design compute lambda = N k $V^{2}$ and read the expected published inflation off the exact curve. The curse is large, two- to tenfold, when lambda is small (the barely-significant designs), decays to one as lambda grows, and is larger at higher degrees of freedom for fixed lambda.

\begin{figure}[htbp]\centering
\includegraphics[width=\linewidth]{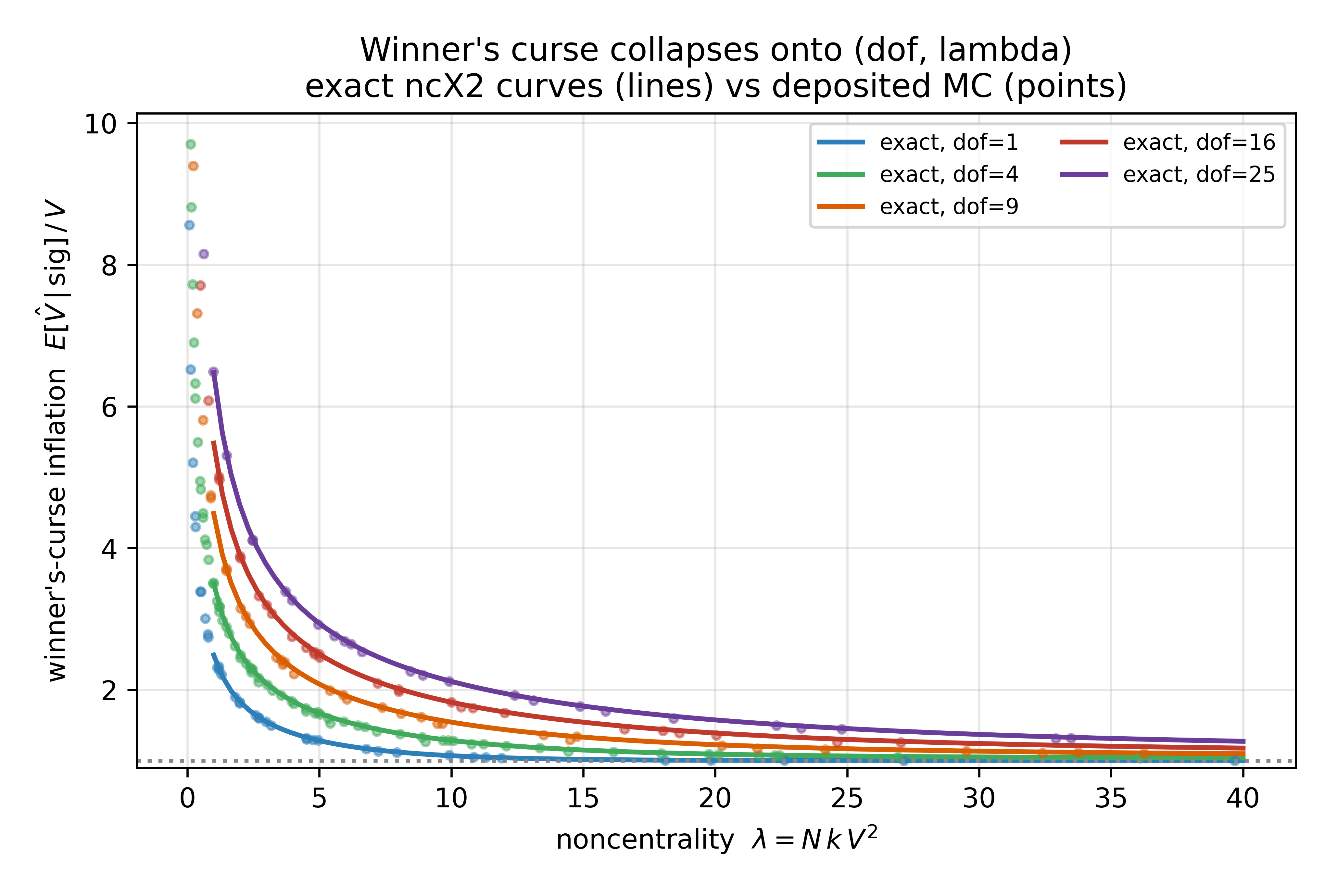}
\caption{\textbf{Figure 4.} The winner's curse is one surface. Significance-conditional inflation E[V given p < .05] / V against the noncentrality lambda = N k $V^{2}$: exact noncentral-chi-square curves by degrees of freedom (lines), with the deposited Monte-Carlo grid (points) lying on them. The whole shape-by-N-by-V grid collapses onto (dof, lambda). Generated by rerun/m0e\_curse\_exact.py, with the Monte-Carlo points from rerun/m0e\_curse.csv.}
\end{figure}

\section{Prevalence}

Every two-way table formed from pairs of categorical variables in the public Rdatasets collection was retained if it had at least as many observations as cells (N >= R*C), N >= 30, and was not degenerate (naive V < 0.999, dropping the deterministic corner): 4,129 tables from 291 datasets. The filter is applied to the raw scan by the deposited script make\_analyzable.py.

\begin{table}[htbp]\centering\small
\caption{\textbf{Table 3.} What the reported effect sizes describe. Values computed by the deposited script scan\_effect\_sizes.py.}
\begin{tabularx}{\textwidth}{>{\raggedright\arraybackslash}X>{\raggedright\arraybackslash}X>{\raggedright\arraybackslash}X}
\hline
 & count & share \\
\hline
analyzable two-way tables & 4,129 &  \\
carrying a Cohen label (small / medium / large) & 2,583 &  \\
... on tables with no significant association at all & 564 & 21.8\% \\
... reading medium or large, reclassified downward by Bergsma's correction, with no significant association & 101 & 30 datasets \\
falling to negligible under Bergsma's correction & 429 &  \\
\hline
\end{tabularx}
\end{table}

The 21.8\% is robust to the inclusion rule (15.5\% to 31.3\% across alternative thresholds; 21.1\% if the degeneracy screen is removed). Tables from the same dataset are not independent: of the 240 datasets that report any labeled effect, 115 (48\%) carry at least one label on a non-significant table, but the median dataset contributes a zero share and the twenty most prolific datasets supply 61\% of the 564, so the pooled rate is driven by a minority of datasets with many cross-tabulations. A cluster bootstrap that resamples whole datasets gives a 95\% interval of 17 to 27\% for the pooled rate. The figure is a property of this corpus of tables, not a survey of published papers. What the row does and does not say is narrow and sufficient: "no significant association" is absence of evidence, not proof of no association, and the claim is only that an effect size was given an interpretive label on a table that provided no evidence of any effect at all.

\begin{figure}[htbp]\centering
\includegraphics[width=\linewidth]{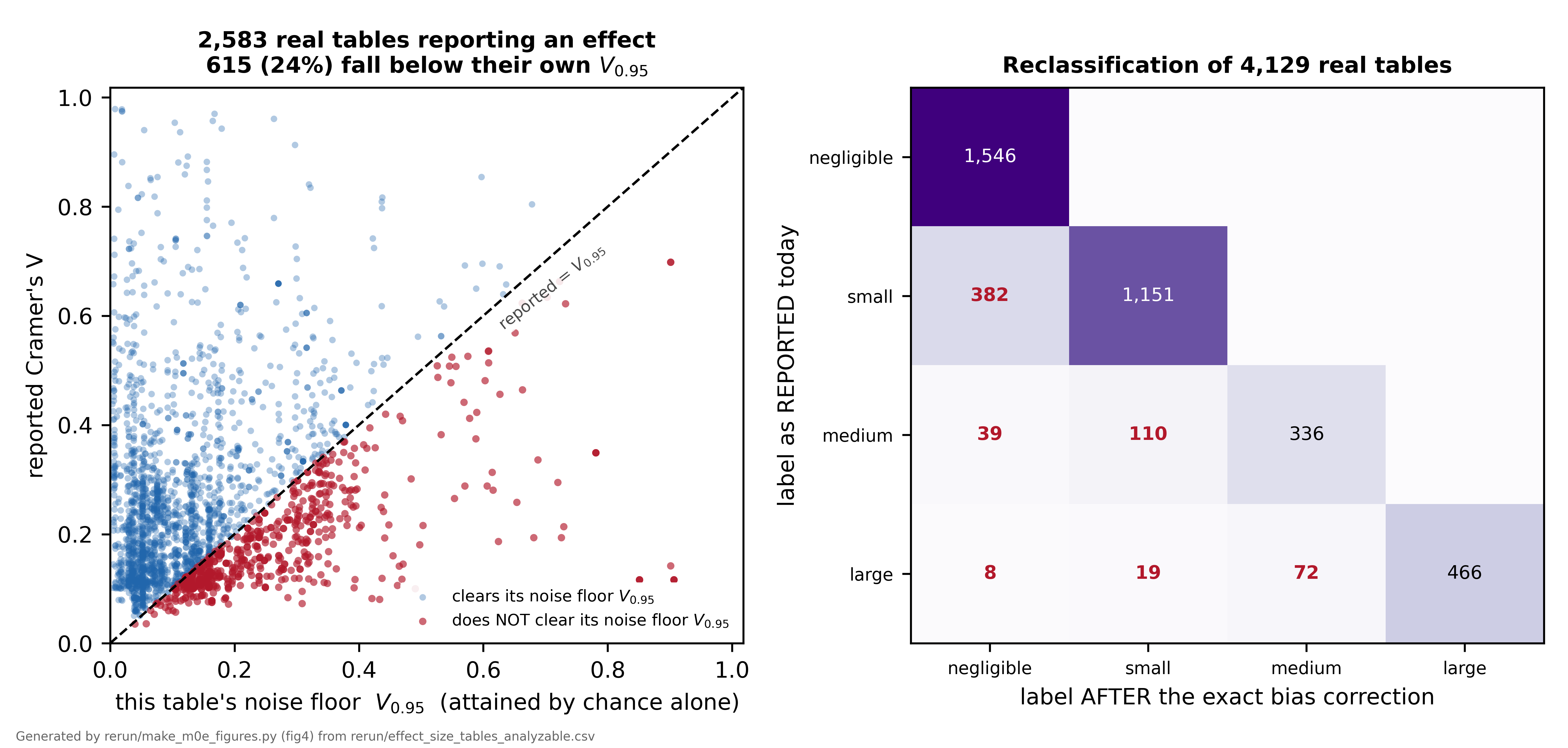}
\caption{\textbf{Figure 5.} 4,129 real two-way tables: reported Cramer's V against each table's own noise floor $V_{0}.95$ (left), and the reclassification produced by Bergsma's correction (right). Generated by rerun/make\_m0e\_figures.py from rerun/effect\_size\_tables\_analyzable.csv.}
\end{figure}

\section{The failure mode occurs in print}

The prevalence scan measures a corpus of tables, not a corpus of papers. To establish that the failure mode reaches publication we audited open-access papers that report a chi-square-family effect size with enough information to reconstruct the check (R, C, N, and the reported value), drawing deliberately from four publishers so that the result is not an artifact of one journal family: Frontiers, MDPI, the Public Library of Science (PLOS), and BioMed Central (BMC). Thirty papers reported enough to check (Frontiers 13, MDPI 8, PLOS 5, BMC 4). For every reported table we verified internal consistency (the reported chi-square reproduces $V^{2}$ N k to rounding) and recomputed the noise floor independently.

This is an existence proof, not a prevalence estimate: the search was steered toward the phenomenon, so the sample is selected on the outcome and cannot support a claim about how often it occurs. What the broadened search does establish is that the failure is not confined to one journal family. Of 90 two-way rows with all four quantities recoverable, 50 report a value below their own noise floor, across 19 of the audited papers and all four publishers; 29 of those below-floor rows, across 10 papers, carry an explicit magnitude label. The mirror-image error on the significance side, reading a non-significant result as proof of no effect, is itself common in print (Murphy et al. 2025 find it in 76 to 85\% of psychology articles that discuss a non-significant finding); our concern is the effect-size side.

The mislabeling is not confined to Cohen's small band; it reaches the medium band, where the word claims a substantively meaningful effect. In Clintberg and colleagues (2026, Frontiers in Child and Adolescent Psychiatry), a longitudinal cohort of infant siblings later diagnosed as autistic is split into three earliest-diagnosis groups and cross-tabulated against the presence of parental concern across ten developmental domains; at the nine- and twelve-month reports none of the domain tests reaches significance, yet the paper calls a run of them "moderate," including a 3 x 2 sleep-domain table at N = 38 with Cramer's V = 0.38 against a floor of 0.40 and a language-regression table at N = 61 with V = 0.31 against a floor of 0.31. These are values sitting squarely in Cohen's medium band, attached to tables that supply no evidence of any association. The same error reaches the review literature: Blessin and colleagues (2022, IJERPH), synthesizing 221 resilience trials, report that the mode of intervention delivery does not differ between Western and Eastern countries (chi-square not significant, p = 0.086) while labeling the accompanying Cramer's V of 0.25 "moderate," where the floor for that 9 x 2 table is 0.27. Across the audit, seven below-floor rows in three papers carry a moderate or medium label.

The small-band cases are worth naming too, because they span designs, fields, and publishers. In Ashrafi and colleagues (2025, BMC Psychiatry), six non-significant contingency tables relating fecal-incontinence status to demographic and pharmacological factors in 200 bipolar-disorder inpatients are each labeled "Small"; the starkest is a 2 x 2 with Cramer's V = 0.029 (p = 0.68) against a floor of 0.14, a magnitude word on a value a fifth of what chance alone produces. In Barbu and colleagues (2026, Healthcare), a 2 x 6 comparison of BMI category by menopausal status in 171 women reports V = 0.15 (p = 0.54) and calls it "a small effect size", where the floor is 0.25. In Hauger and colleagues (2025, PLOS ONE), among 130 rehabilitation patients a 2 x 5 table reports V = 0.22 and a 2 x 2 reports V = 0.08, both labeled a "low association", where the floors are 0.27 and 0.17. And in Yakushina, Chichinina and Dolgikh (2025, Frontiers in Psychology), a 2 x 2 at N = 88 has $V_{0}.95$ = 0.209 while the paper reports six values of V between 0.12 and 0.18 on a scale calling 0.10 < V < 0.30 a small effect, so two thirds of Cohen's small band lies below the noise floor at that design.

The honest cases are as instructive as the mislabeled ones. In Tilp and colleagues (2026), a volleyball study, nine of ten reported values of Cramer's V fall below their floors (2 x 2 and 2 x 3 tables at N = 85 to 175), but the paper attaches no verbal label, which is the better practice; the residual issue is only that a bare magnitude is still printed beside each non-significant table. In Avugos and Haleva (2026), six one-way goodness-of-fit tests of NBA birth quartiles all fall below their floors while the substantive conclusion, that there is no relative-age effect, is correct, the error confined to the effect-size sentence ("Effect sizes were small"). And Francisco and colleagues (2020) already print a dash instead of V for every non-significant chi-square, an existing-practice instance of the remedy proposed here.

The audit also shows why the floor cannot be replaced by a fixed threshold: a V of 0.059 at N = 2,226 clears its floor while a V of 0.18 at N = 88 does not, and the same nominal V of 0.13 is below floor at N = 128 and above it at N = 1,051. A bare V is uninterpretable without the shape and the sample size, which is precisely the argument. The full audit, with every reported statistic, the recomputed floor, and the internal-consistency check, is deposited as literature\_audit.csv.

The remedy is a report, and it runs in a browser. The deposited tool takes a pasted table and returns the reference, the noise floor, and the supportability verdict, so the floor travels with the number instead of a reader having to compute it. Figure 7 shows that report on a public example beside what software prints today; the print failures catalogued above are exactly what it is built to prevent.

\begin{figure}[htbp]\centering
\includegraphics[width=\linewidth]{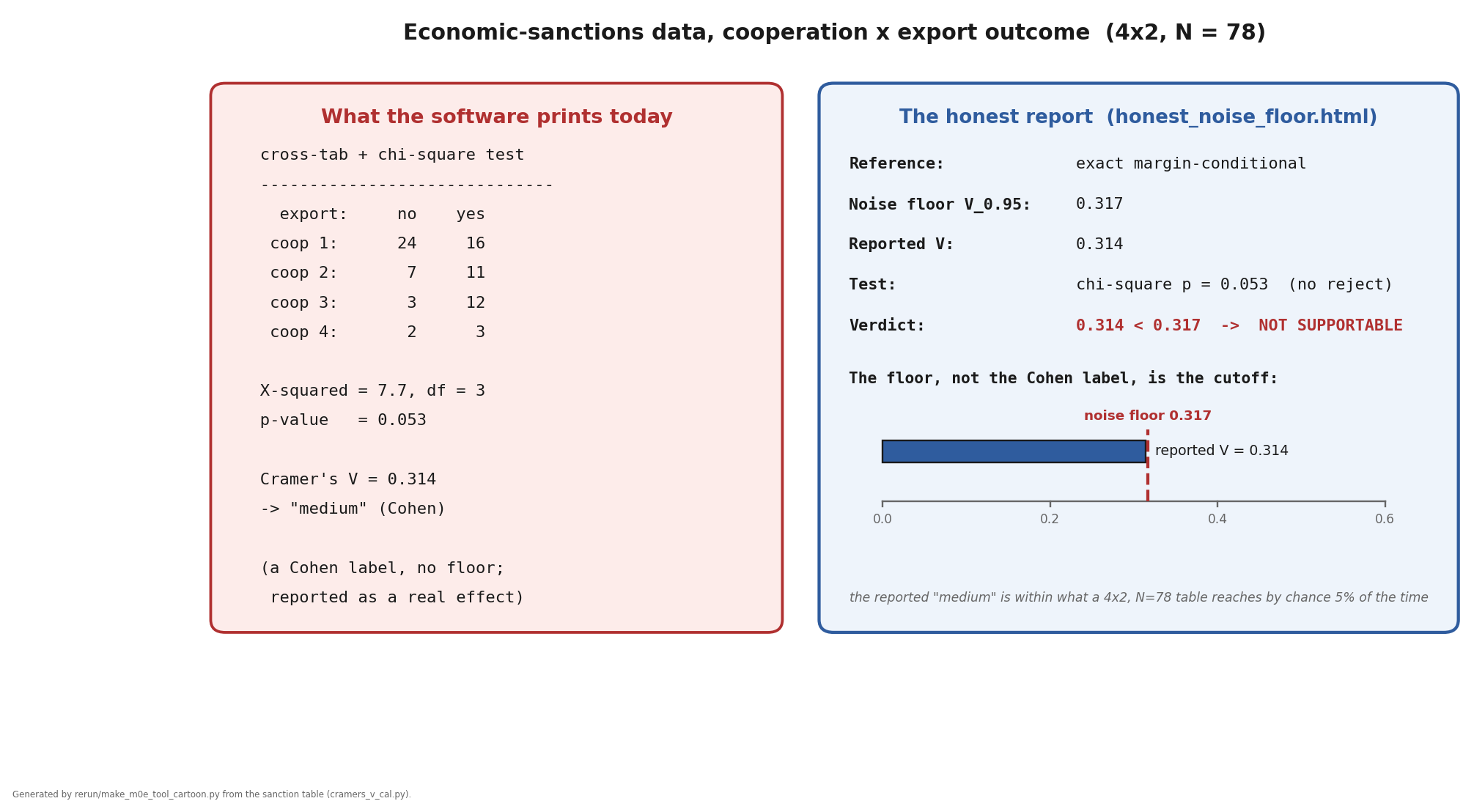}
\caption{\textbf{Figure 7.} The honest report the deposited browser tool produces (right) beside what general-purpose software prints today (left), for the economic-sanctions data (cooperation $\times$ export outcome, 4$\times$2, N = 78). Software prints a bare Cramér's V = 0.314 and reads it against a "medium" Cohen label; the report adds the reference, the per-table noise floor $V_{0}.95$ = 0.317, and the verdict — the reported V sits just below its own floor and the chi-square test does not reject (p = 0.053), so the "medium" magnitude is within what this design reaches by chance and is not supportable. Generated by rerun/make\_m0e\_tool\_cartoon.py.}
\end{figure}

\section{Relation to prior work}

\begin{table}[htbp]\centering\small
\caption{\textbf{Table 4.} Established components and the contribution.}
\begin{tabularx}{\textwidth}{>{\raggedright\arraybackslash}X>{\raggedright\arraybackslash}X>{\raggedright\arraybackslash}X>{\raggedright\arraybackslash}X}
\hline
Strand & Prior art & What this adds & Novelty \\
\hline
The family as one object & Pearson 1900; Cramer 1946; Tschuprow 1925; Cohen 1988 & Organization only & None \\
Exact null mean $E[\phi ^{2}]$ = dof/(N-1) & Tschuprow 1925; Bartlett 1937 & Numerical verification & None \\
Bias correction & Bergsma 2013 & Used unchanged & None \\
Reporting the critical effect size beside the observed one & Cohen 1988; Bloom 1995; Hoenig and Heisey 2001; Perugini et al. 2025 & Specialized to the chi-square family, on the effect-size scale, as a per-table gate & Low \\
Winner's curse / Type M error & Ioannidis 2008; Gelman and Carlin 2014 & Magnitudes in this family; both defects driven by the same $X^{2}$ & Moderate \\
\textbf{The exact margin-conditional floor} & asymptotic critical value; exact engine (Dwyer 2026a) & A per-table gate valid where the asymptotic floor miscalibrates (0.10 vs 0.05 false-alarm; Figure 2) & \textbf{High} \\
\textbf{Prevalence in real tables} & -- & 4,129 tables; 21.8\% of labeled effects on non-significant tables (cluster-robust 17-27\%) & \textbf{High} \\
\hline
\end{tabularx}
\end{table}

We did not discover that Cramer's V is biased upward, we did not derive the exact null mean, we did not invent the bias correction or the winner's curse, and we did not originate the idea of reporting the critical effect size. What we claim is an exact per-table reporting gate for this family, valid on the sparse and heterogeneous tables where the asymptotic version and the general recommendation are miscalibrated, and a measurement of how often a labeled effect in real data would be changed by it. A fuller account is given in the companion novelty review.

Blessin and colleagues (2022) make the gap concrete. Theirs is a careful systematic review; the authors computed Cramer's V, ran the chi-square, and reported the non-significant p, so every quantity needed to catch the mislabel was already on the page. What no step in that workflow supplied, and what this paper adds, is the one comparison that settles it: the reported V of 0.25 against the 0.27 the same 9 x 2 table at N = 221 produces by chance. A fixed Cohen threshold cannot make that comparison, because 0.25 sits in the small-to-medium range at every shape and every sample size; the value is below floor only by virtue of this table's degrees of freedom and total, and on sparser tables the defensible floor is the exact margin-conditional one rather than the asymptotic approximation. This is not a repudiation of Cohen's benchmarks, which remain the right instrument at the design stage (Section 9), where there is as yet no estimate to gate; the claim is only that at the reporting stage a fixed label is no substitute for the table's own floor. Reporting an effect size, applying a label, and noting significance are all established practice; printing the per-table floor beside the value is not, and it is what converts the numbers already in hand into a correct decision.

\section{What we are and are not asserting}

A claim of the form "this table has no association" would be an absence-of-evidence fallacy, especially poor in a paper about statistical hygiene. We do not make it anywhere.

\textbf{The claim, in full.} A magnitude label was attached to a number that does not exceed the value the same statistic attains on a table with no association at all. That is a statement about the reporting, not about the world.

A table below its floor may well contain a real association the study was too small to detect; indeed that is often the more likely reading, and it does not weaken the point, since the reported magnitude was not supportable either way. We therefore never write "no effect" of a real table. Where a true null is asserted (Figures 1 and 2 use simulated data), the null is known because we generated it. The papers cited in Section 7 are cited as instances of a reporting convention, never as accusations; in every case the authors' substantive conclusion was correct and the error was in the effect-size sentence.

\#\#\# Why a label is not enough

A natural objection is that we attack a straw man: nobody believes Cohen's labels are precise, and reporting the number and the floor is compatible with also attaching a word. But a label discards the two things a reader needs. It discards uncertainty: a "medium" effect from N = 20 and one from N = 20,000 receive the same word, though the first may be consistent with zero. And it discards magnitude: V = 0.31 and V = 0.49 are both "medium," while V = 0.29 and V = 0.31 fall on opposite sides of a cliff. It is the significance dichotomy in effect-size clothing, three bins in place of two, and it is stamped on a number that is biased upward and can fire below the noise floor. Figure 6 makes this exact: Cohen's bands are identical at every N because the conventions ignore the sample size, while the floor rises as N falls, so on a 3 x 3 table at N = 50 the floor reaches $V_{0}.95$ = 0.31 and the whole "small" band and the base of "medium" describe sampling variation.

None of this retires Cohen's conventions from the one place they were built for: a-priori design. Before a study is run there is no estimate to label and no interval to draw, and a fixed, shared, deliberately crude benchmark is the right instrument for choosing a sample size; Cohen offered the labels "for use only when no better basis is available." At the design stage the label is a planning yardstick and $V_{0}.95$ is its natural companion, itself a design quantity. At the reporting stage the number, the floor, and the interval carry everything the label was reaching for plus the uncertainty it omits, so the word becomes at most a caption, gated by the floor. We therefore recommend retiring the label as the primary report of a result, and retaining Cohen's conventions in the power calculation that precedes the study.

\begin{figure}[htbp]\centering
\includegraphics[width=\linewidth]{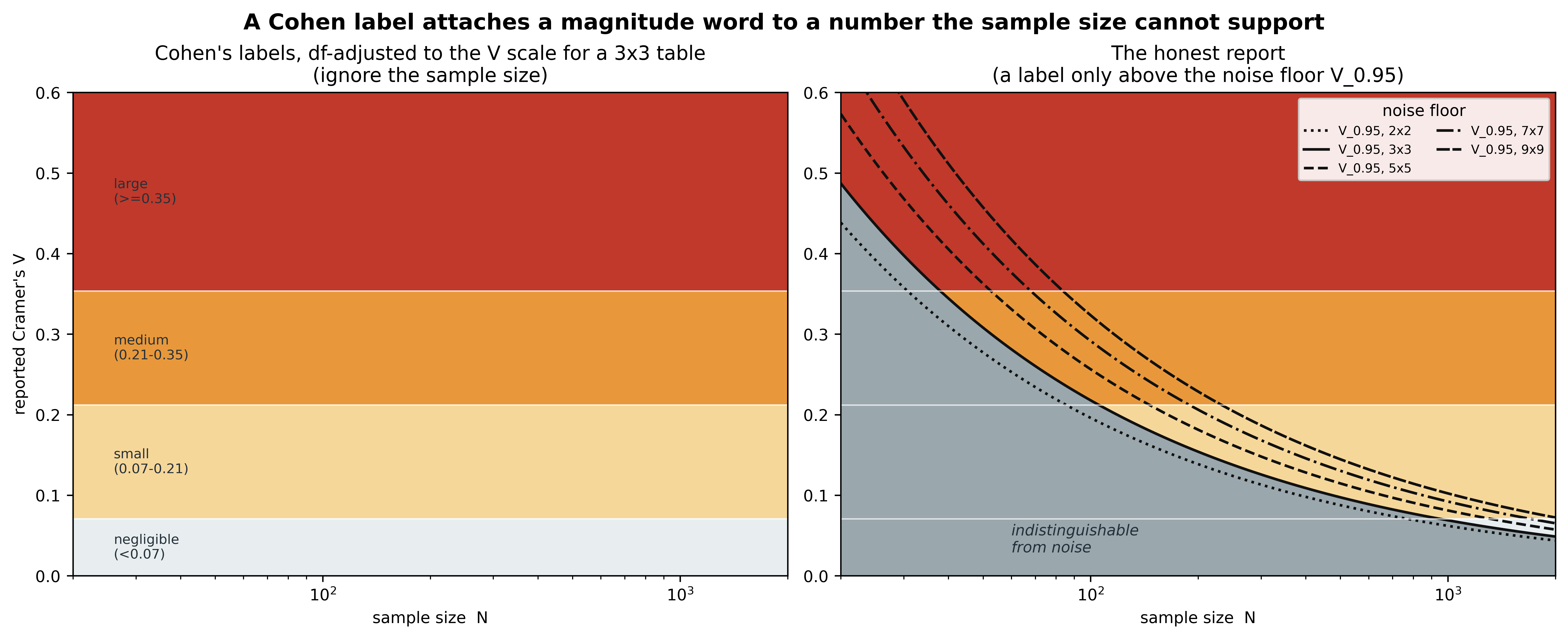}
\caption{\textbf{Figure 6.} Cohen's last-resort labels versus the honest, noise-floor-gated report, over the sample size N and the reported Cramer's V. Left: Cohen's bands, mapped to the V scale for a 3 x 3 table (V = w / sqrt(k); df-adjusted edges 0.07, 0.21, 0.35), identical at every N. Right: the same bands, with every cell below the design's noise floor $V_{0}.95$(N) repainted as indistinguishable from noise (floor curves for 2x2, 3x3, 5x5, 7x7, 9x9, stacked lowest to highest). The gray region is gated by the 3x3 floor, which reaches 0.31 at N = 50, while the 9x9 floor there reaches 0.46, so on the larger table the whole small and medium bands and the base of "large" describe sampling variation. On a 2x2 V equals w and the edges revert to Cohen's originals; the larger the table, the higher its floor climbs, so the shaded below-floor area is conservative for every table above 3x3. The exact margin-conditional floor is the companion engine's (Dwyer 2026a). Rendered by the deposited script make\_cohen\_vs\_honest.py.}
\end{figure}

\section{Scope and limits}

$V_{0}.95$ is a monotone transform of the critical value and inherits the properties of the underlying test: it is a statement about the design, not about the truth of the null. Three floors are in play and we distinguish them. The \textbf{asymptotic} floor is the elementary central-chi-square evaluation; it is exact on dense balanced tables and miscalibrates otherwise (Figure 2). The \textbf{exact-moment} floor shipped with this paper is computed from the exact margin-conditional moments and agrees with the asymptotic form to within 0.7\% on the balanced designs examined, but was itself calibrated on balanced margins and, under strongly heterogeneous margins with minimum expected count below about 0.05, its false-alarm rate ranges from 0.016 to 0.066. The \textbf{exact margin-conditional} floor of the companion engine (Dwyer 2026a) is the defensible one in that corner and supersedes the exact-moment refinement; analyses there are flagged and carried by the accompanying exact test rather than pooled. We do not claim the floor is trustworthy where it is not.

Cohen's thresholds are conventions; the findings do not depend on them, only on the distance between a reported value and the value the same statistic reports on nothing. The prevalence scan covers a dataset collection, not a sample of published papers, and its tables are not independent. The literature audit of Section 7 is an existence proof only, selected on the outcome and skewed toward open access by retrieval; a genuine prevalence figure for the published literature would require a pre-registered random sample from a defined frame, screened blind to outcome, which we do not claim.

\section{Reproducibility and companions}

All results are reproducible from the deposited code (rerun/effect\_sizes.py, es\_moments.py, m0e\_winners\_curse.py, m0e\_curse\_exact.py, scan\_effect\_sizes.py, make\_analyzable.py, make\_m0e\_figures.py, make\_m0e\_floor\_calibration.py, make\_cohen\_vs\_honest.py). The derivation companion carries the full treatment, including the withdrawn claims and why they were withdrawn. The literature audit is deposited as M0e\_Literature\_Audit.md with its raw data (rerun/literature\_audit.csv). An independent adversarial methods review is deposited as M0e\_Methods\_Review.md. A browser tool (honest\_effect\_size.html) computes the report for any pasted table. The exact-computation engine is the companion Dwyer (2026a); the point estimate and confidence interval are the companion Dwyer (2026b).

\section{Conclusion}

The mathematics has been available since 1937 and the correction since 2013, yet labeled effects still land on tables with no detectable signal, in about a fifth of the labeled tables in our corpus. A plausible part of the obstacle is presentation: the critical value is reported in units nobody interprets, and on the tables where it matters most the elementary version of it is wrong. Report the exact noise floor $V_{0}.95$ in the units the reader actually interprets, next to the number they are about to name, and the problem becomes self-evident. We recommend that no effect size for a contingency table be reported without its interval and its noise floor $V_{0}.95$, that the floor be computed in exact margin-conditional form on sparse or heterogeneous tables, and that it be computed at the design stage, where it states the smallest effect the planned study could ever honestly report.

\section*{Declarations}

\textbf{Ethics approval and consent to participate.} Not applicable; all data are public and de-identified.

\textbf{Consent for publication.} Not applicable.

\textbf{Clinical trial number.} Not applicable.

Availability of data and materials. All code and derived data are deposited in a permanent, version-controlled archive (Zenodo; concept DOI 10.5281/zenodo.21664321, which resolves to the latest version; MIT for code, CC BY 4.0 for documents and data). The mathematics is consolidated in the derivation companion M0e\_Derivations.md.

\textbf{Disclosure of interest.} The author develops and hosts the open-source software and associated web domains (the trialdesign.com applications) that implement methods discussed in this work; no other competing interests are declared.

\textbf{Funding.} None.

\textbf{Authors' contributions.} WJD is the sole author and is responsible for the conception, analysis, software, and writing of this work.

\textbf{Use of generative AI.} In preparing this manuscript the author used a generative-AI assistant (Claude, Anthropic; Opus 4) for drafting and editing prose, generating figure code, and constructing and formatting tables. All AI-assisted output was reviewed and verified by the author; every reported figure and number regenerates deterministically from the openly deposited code, and the author takes full responsibility for the content of this work.

\section*{References}

\noindent Ashrafi S, Hedayati A, Karimi Kordestani B, Najafzadeh H. Fecal incontinence in hospitalized bipolar disorder patients: prevalence, gender disparities, and associations with demographic and pharmacological factors. BMC Psychiatry. 2025. doi:10.1186/s12888-025-07650-1.\par\smallskip

\noindent Avugos S, Haleva Y. Relative age effect in high-skill labor markets: evidence from the NBA. Frontiers in Sports and Active Living. 2026;8:1787778.\par\smallskip

\noindent Barbu CG, Nistor C, Albu A, Martin S, Oprea TE, Sirbu AE, Vlad A, Fica S. Fat mass is associated with aging rather than menopausal transition. Healthcare. 2026;14(3):333.\par\smallskip

\noindent Bartlett MS. Properties of sufficiency and statistical tests. Proceedings of the Royal Society A. 1937;160:268-282.\par\smallskip

\noindent Bergsma W. A bias correction for Cramer's V and Tschuprow's T. Journal of the Korean Statistical Society. 2013;42(3):323-328.\par\smallskip

\noindent Blessin M, Lehmann S, Kunzler AM, van Dick R, Lieb K. Resilience interventions conducted in Western and Eastern countries: a systematic review. International Journal of Environmental Research and Public Health. 2022;19(11):6913.\par\smallskip

\noindent Bloom HS. Minimum detectable effects: a simple way to report the statistical power of experimental designs. Evaluation Review. 1995;19(5):547-556.\par\smallskip

\noindent Clintberg K, Sacrey L-AR, Zwaigenbaum L, Brian JA, Smith IM, et al. Parental concerns correspond to earliest age of autism diagnosis in increased likelihood infant cohort. Frontiers in Child and Adolescent Psychiatry. 2026;4:1722543.\par\smallskip

\noindent Cohen J. Statistical Power Analysis for the Behavioral Sciences. 2nd ed. Hillsdale, NJ: Erlbaum; 1988.\par\smallskip

\noindent Cramer H. Mathematical Methods of Statistics. Princeton: Princeton University Press; 1946.\par\smallskip

\noindent Dwyer WJ. Exact conditional distributions of chi-square-family statistics for two-way contingency tables, by cell-separable dynamic programming. Companion manuscript, submitted for publication; 2026a.\par\smallskip

\noindent Dwyer WJ. An honest effect size for contingency tables: nothing can be unbiased, so the error must be placed. Companion manuscript, submitted for publication; 2026b.\par\smallskip

\noindent Fisher RA. The general sampling distribution of the multiple correlation coefficient. Proceedings of the Royal Society A. 1928;121:654-673.\par\smallskip

\noindent Francisco R, Pedro M, Delvecchio E, Espada JP, Morales A, Mazzeschi C, Orgiles M. Psychological symptoms and behavioral changes in children and adolescents during the early phase of COVID-19 quarantine in three European countries. Frontiers in Psychiatry. 2020;11:570164.\par\smallskip

\noindent Gelman A, Carlin J. Beyond power calculations: assessing Type S (sign) and Type M (magnitude) errors. Perspectives on Psychological Science. 2014;9(6):641-651.\par\smallskip

\noindent Hauger SL, Sundet AS, Lovstad M, Havnes IA, Andelic N, Ustvedt C, Manum G, Hoye H, et al. Rates and temporal onset of mental health disorders during inpatient rehabilitation after acute physical injury or illness: an observational cohort study. PLOS ONE. 2025. doi:10.1371/journal.pone.0338207.\par\smallskip

\noindent Hoenig JM, Heisey DM. The abuse of power: the pervasive fallacy of power calculations for data analysis. The American Statistician. 2001;55(1):19-24.\par\smallskip

\noindent Ioannidis JPA. Why most discovered true associations are inflated. Epidemiology. 2008;19(5):640-648.\par\smallskip

\noindent Lakens D, Scheel AM, Isager PM. Equivalence testing for psychological research: a tutorial. Advances in Methods and Practices in Psychological Science. 2018;1(2):259-269.\par\smallskip

\noindent Murphy J, Merz S, Reimann M, Fernandez R. Nonsignificance misinterpreted as an effect's absence in psychology: prevalence and temporal analyses. Royal Society Open Science. 2025;12(3):242167.\par\smallskip

\noindent Pal T, Iantovics LB, et al. Risk factors for cognitive dysfunction amongst patients with cardiovascular diseases. Frontiers in Public Health. 2024;12:1385089.\par\smallskip

\noindent Pearson K. On the criterion that a given system of deviations from the probable in the case of a correlated system of variables is such that it can be reasonably supposed to have arisen from random sampling. Philosophical Magazine. 1900;50:157-175.\par\smallskip

\noindent Perugini A, Gambarota F, Toffalini E, Lakens D, Pastore M, Finos L, Altoe G. The benefits of reporting critical-effect-size values. Advances in Methods and Practices in Psychological Science. 2025;8(2). doi:10.1177/25152459251335298.\par\smallskip

\noindent Smithson M. Confidence Intervals. Thousand Oaks: Sage; 2003.\par\smallskip

\noindent Tilp M, Pusch C, Medeiros AIA, Wieland B, Prosch Y, Zentgraf K, Kneubuhl I, Giatsis G. Association between spike technique and injuries in competitive volleyball players. Frontiers in Sports and Active Living. 2026;8:1737436.\par\smallskip

\noindent Tschuprow AA. Grundbegriffe und Grundprobleme der Korrelationstheorie. Leipzig: Teubner; 1925.\par\smallskip

\noindent Yakushina A, Chichinina E, Dolgikh A. Chess classes and executive function skills in 5-6 year old children. Frontiers in Psychology. 2025;16:1564963.\par\smallskip

\end{document}